\PassOptionsToPackage{unicode}{hyperref}
\PassOptionsToPackage{hyphens}{url}
\documentclass[
  letterpaper,
  10 pt,
  conference]{ieeeconf}
\usepackage{amssymb,amsmath}
\usepackage{ifxetex,ifluatex}
\ifnum 0\ifxetex 1\fi\ifluatex 1\fi=0 
  \usepackage[T1]{fontenc}
  \usepackage[utf8]{inputenc}
  \usepackage{textcomp} 
\else 
  \usepackage{unicode-math}
  \defaultfontfeatures{Scale=MatchLowercase}
  \defaultfontfeatures[\rmfamily]{Ligatures=TeX,Scale=1}
\fi
\IfFileExists{upquote.sty}{\usepackage{upquote}}{}
\IfFileExists{xurl.sty}{\usepackage{xurl}}{} 
\IfFileExists{bookmark.sty}{\usepackage{bookmark}}{\usepackage{hyperref}}
\hypersetup{
  pdftitle={Resilient Self/Event-Triggered Consensus Based on Ternary Control},
  pdfauthor={Hiroki Matsume, Yuan Wang, and Hideaki Ishii},
  hidelinks,
  pdfcreator={LaTeX via pandoc}}
\usepackage{graphicx}
\makeatletter
\def\maxwidth{\ifdim\Gin@nat@width>\linewidth\linewidth\else\Gin@nat@width\fi}
\def\maxheight{\ifdim\Gin@nat@height>\textheight\textheight\else\Gin@nat@height\fi}
\makeatother
\setkeys{Gin}{width=\maxwidth,height=\maxheight,keepaspectratio}
\makeatletter
\def\fps@figure{htbp}
\makeatother
\providecommand{\tightlist}{%
  \setlength{\itemsep}{0pt}\setlength{\parskip}{0pt}}
\usepackage[nobreak]{cite}
\usepackage{algorithm}
\usepackage{algpseudocode}

\newcommand{\qed}{\hfill\square}
\newenvironment{cslreferences}%
  {}%
  {\par}
\renewenvironment{cslreferences}{\everypar{\setlength{\hangindent}{2em}}}{\par}
\makeatletter
\@ifpackageloaded{subfig}{}{\usepackage{subfig}}
\@ifpackageloaded{caption}{}{\usepackage{caption}}
\@ifpackageloaded{float}{}{\usepackage{float}}
\floatstyle{ruled}
\@ifundefined{c@chapter}{\newfloat{codelisting}{h}{lop}}{\newfloat{codelisting}{h}{lop}[chapter]}
\floatname{codelisting}{Listing}

\@ifpackageloaded{cleveref}{}{\usepackage{cleveref}}
\crefname{figure}{Fig.}{Fig.}
\Crefname{figure}{Fig.}{Fig.}
\crefname{table}{TABLE}{TABLE}
\Crefname{table}{TABLE}{TABLE}
\crefname{equation}{}{}
\Crefname{equation}{}{}
\crefname{listing}{lst.}{lsts.}
\Crefname{listing}{Lst.}{Lsts.}
\crefname{section}{sec.}{secs.}
\Crefname{section}{Sec.}{Secs.}
\crefname{codelisting}{\cref@listing@name}{\cref@listing@name@plural}
\Crefname{codelisting}{\Cref@listing@name}{\Cref@listing@name@plural}
\makeatother
\newlength{\cslhangindent}
\title{\LARGE \bf Resilient Self/Event-Triggered Consensus Based on
Ternary Control}
\author{Hiroki Matsume, Yuan Wang, and Hideaki Ishii \thanks{H. Matsume,
Y. Wang, and H. Ishii are with the Department of Computer Science, Tokyo
Institute of Technology, Yokohama, 226-852, Japan. E-mails:
\tt{\small{matsume.h.aa@m.titech.ac.jp}}, \tt{\small{wang.y.bb@m.titech.ac.jp}}, \tt{{\small ishii@c.titech.ac.jp}}}\thanks{This
work was supported in part by the JST CREST Grant No.~JPMJCR15K3 and by
the JSPS KAKENHI Grant No.~18H01460. The financial support provided by
the China Scholarship Council is also acknowledged.}}
\date{May 2020}

\IEEEoverridecommandlockouts

\begin{document}

\maketitle
\thispagestyle{empty}
\pagestyle{empty}

\begin{abstract}
The paper considers the problem of multi-agent consensus in the presence
of adversarial agents which may try to prevent and introduce undesired
influence on the coordination among the regular agents. To our setting,
we extend the so-called mean subsequence reduced algorithms with the aim
to reduce the amount of communication via two measures: The agents
exchange information in the form of ternary data at each transmission
and moreover keep the frequency of data exchange low by employing self-
and event-triggered communication. We will observe that in hostile
environments with adversaries, the self-triggered approach can bring
certain advantages over the event-triggered counterpart.
\end{abstract}

\hypertarget{introduction}{%
\section{INTRODUCTION}\label{introduction}}

The study of multi-agent systems for distributed coordination has
received a lot of attention in the field of systems control. In such
systems, a number of agents form a network over which they communicate
with each other using wired or wireless channels and individually make
decisions to achieve their global objectives {[}1{]}. The increase in
networking has brought about new critical issues related to
cybersecurity to multi-agent systems {[}2{]}, {[}3{]}. Even if the
adversarial attacks are local and limited to a small portion of the
agents, misbehaviors of them can potentially harm the overall system
performance. Thus, development of methods robust to such attacks has
become of great importance.

Resilience in consensus-type algorithms against malicious adversaries
has been long studied in the area of fault-tolerant distributed
algorithms in computer science {[}4{]}. Our approach follows this line
of research, which has recently become active in the systems control
community. The basic problem setting is as follows: The agent network is
known to contain up to \(F\) adversarial agents whose identities are
unknown to the nonfaulty, regular agents. The adversarial agents may
harm the consensus process as they can transmit false data to their
neighbors to be used in their updates. The result can be that consensus
takes place at dangerous values or the agents may even be split into
disconnected groups.

For resilient consensus, the class of mean subsequence reduction (MSR)
algorithms is effective to mitigate the influence of adversaries. With
the knowledge of the bound \(F\) on the number of adversaries, the
agents will disregard neighbors whose states take especially large or
small values. This type of algorithms has been applied to consensus
problems with agents' states taking real values (e.g., {[}5{]}) though
most of the past studies have been limited to networks of complete
graphs; however, the use of such networks is very limited in practice.
More recently, noncomplete graph cases have been addressed and
topological conditions in terms of robust graphs have been obtained
{[}6{]}, {[}7{]}. The MSR approach for multi-agent consensus has been
extended to quantized states {[}8{]}, agents with higher order dynamics
{[}9{]}, {[}10{]}, and applications to robotic networks {[}11{]} and
clock synchronization of wireless sensor networks {[}12{]}.

In this paper, we place special emphases on the reduction of
communication load for the interactions among agents. This is to be
achieved by two measures: One is by transmissions of very coarse data by
quantization using only ternary outputs {[}13{]}, {[}14{]}. The simple
data expression is helpful not only in keeping the information minimal
in each data sent, but also in reducing the number of transmissions; if
the data to be sent is the same as the previous time, this communication
can be accomplished implicitly.

The other is to reduce the transmission frequency via event-based
communication techniques (e.g., {[}15{]},{[}16{]}). In particular, we
employ the self-triggered control approach {[}14{]}, {[}16{]}--{[}18{]},
under which the agents determine their next update and transmission
times dynamically according to the values received from other agents and
its own. This enables the system not to respond to data received before
the next update instant comes. It turns out that this feature is useful
in hostile environments, where adversaries may make unnecessary
communication to disturb others or to jam the communication. By
contrast, in the similar, but more popular approach of event-triggered
control (e.g., {[}16{]}, {[}19{]}, {[}20{]}), agents must make updates
in their values each time data is received; this is because the agents
must check if the current state is sufficiently different from the one
last transmitted, in which case the next transmission will be made.

For comparison reasons, in this paper, we propose resilient consensus
protocols by incorporating the MSR algorithm into both self-triggered
and event-triggered protocols with ternary control. It will be found
that from the theoretical viewpoint, the two protocols can be treated in
a similar manner. Both protocols are capable in achieving resilient
consensus via asynchronous update even when time delays are present in
communication {[}21{]}. Furthermore, in both protocols, the control
input for each agent is quantized into three values of \(-1\), \(0\),
and \(1\) {[}14{]}. From the perspective of quantized consensus {[}8{]},
{[}22{]}--{[}24{]}, it is notable that many existing consensus
algorithms rely on randomization-based techniques in the update rules;
these are needed for completing consensus in a finite number of updates
(in a probabilistic sense). We highlight that our approach is
deterministic, which contributes to simplifying the the analysis.

The paper is organized as follows: In Section II, we provide the problem
formulation and introduce the self-triggered protocol with ternary
control. In Section III, we present the MSR-based algorithm and the
result on resilient consensus. Section IV discusses the event-triggered
version of the resilient algorithm. In Section V, we compare the two
event-based algorithms via a numerical example. We conclude the paper in
Section VI.

\hypertarget{problem-formulation}{%
\section{PROBLEM FORMULATION}\label{problem-formulation}}

\hypertarget{preliminaries-on-graphs}{%
\subsection{Preliminaries on graphs}\label{preliminaries-on-graphs}}

Consider the directed graph \(\mathcal G = (\mathcal V, \mathcal E)\)
with the set \(\mathcal V = \{1,2,...,n\}\) of nodes and the set
\(\mathcal E \subset \mathcal V \times \mathcal V\) of edges. The edge
\((j, i) \in \mathcal E\) indicates that node \(j\) can send a message
to node \(i\) and is called an incoming edge of node \(i\). Let
\(\mathcal N_i = \{j: (j, i)\in\mathcal E\}\) be the set of neighbors
and let \(d_i = |\mathcal N_i|\) be the degree of node \(i\). The graph
is said to be connected if for any pair of nodes, there is a sequence of
edges from one to the other. Denote by \(A\) the Laplacian matrix of
\(\mathcal G\).

To establish resilient consensus results, we introduce an important
topological notion of robustness for graphs {[}6{]}.

\textbf{Definition 1.} The graph \(\mathcal G=(\mathcal V, \mathcal E)\)
is called \((r, s)\)-robust (\(r, s<n\)) if for every pair of nonempty,
disjoint subsets \(\mathcal V_1, \mathcal V_2\subseteq \mathcal V\), at
least one of the following holds: (i)
\(\mathcal X_{\mathcal V_1}^r=\mathcal V_1\), (ii)
\(\mathcal X_{\mathcal V_2}^r=\mathcal V_2\), and (iii)
\(|\mathcal X_{\mathcal V_1}^r| + |\mathcal X_{\mathcal V_2}^r|\geq s\),
where \(\mathcal X_{\mathcal{V}_i}^r\) is the set of all nodes in
\(\mathcal V_i\) which have at least \(r\) neighbors outside
\(\mathcal V_i\) for \(i=1, 2\). The graph is said to be \(r\)-robust if
it is \((r, 1)\)-robust.

\begin{figure}[t]
\hypertarget{fig:2robustexample}{%
\centering
\includegraphics[width=\textwidth,height=30mm]{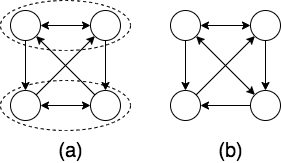}
\caption{Example graphs: (a) not 2-robust and (b)
2-robust}\label{fig:2robustexample}
}
\end{figure}

In \cref{fig:2robustexample}, we display two example graphs with four
nodes. The graph (a) is not 2-robust because any of the nodes in the
subgraphs surrounded by the dotted lines has just one neighbor outside
the subgraph to which it belongs. On the other hand, it can be confirmed
that the graph (b) is 2-robust from the definition.

\hypertarget{self-triggered-consensus-protocol}{%
\subsection{Self-triggered consensus
protocol}\label{self-triggered-consensus-protocol}}

We introduce the self-triggered consensus protocol based on the ternary
controller for the agents to achieve consensus {[}14{]}. In this
protocol, each agent \(i\) has the state variable \(x_i\), the control
\(u_i\) (which is quantized to a ternary value), and the local clock
variable \(\theta_i\). All these variables are defined for time
\(t\geq0\). Each agent triggers update events and transmission events
depending on its own clock.

\hypertarget{dynamics-in-the-hybrid-system-format}{%
\subsubsection{Dynamics in the hybrid system
format}\label{dynamics-in-the-hybrid-system-format}}

We express the self-triggered dynamics for each agent \(i\in\mathcal V\)
as a hybrid system consisting of three variables \(x_i\), \(u_i\), and
\(\theta_i\). At time \(t\), these states satisfy the following
continuous evolution: \begin{equation}
\begin{cases}
\dot x_i = u_i\\
\dot u_i = 0\\
\dot\theta_i = -1
\end{cases}
\label{eq:continuous-evo}\end{equation} except for every \(t\) such that
the timer variable \(\theta_i(t)\) becomes zero. Such time instants are
called the self-triggered times and are denoted by
\(\{t_m^i\}_{m\in\mathbb Z_+}\) with \(t_0^i=0\) and
\(t_m^i<t_{m+1}^i\). The set of all self-triggered times is given by \[
\{t_l\}_{l\in\mathbb Z_+}=\bigcup_{i\in\mathcal{V}}\{t_m^i\}_{m\in\mathbb{Z}_+}.
\] Moreover, let \(\mathcal{U}(t)\) be the set of nodes whose
self-triggered times are equal to \(t\), that is, \[
\mathcal U(t)=\{i\in\mathcal V:\theta_i(t)=0\}.
\]

At such time instants \(\{t_l\}\), agent \(i\)'s states follow the
discrete evolution given by \begin{equation}
\begin{cases}
x_i(t^+) = x_i(t)\\
u_i(t^+) = \begin{cases}
            \mathrm{sign}(f_\varepsilon(\mathrm{ave}_i(t)))&\mathrm{if}\ i\in\mathcal U(t)\\
            u_i(t)&\mathrm{otherwise}
            \end{cases}\\
\theta_i(t^+) = \begin{cases}
            \max\{|\mathrm{ave}_i(t)|,\varepsilon\}&\mathrm{if}\ i\in\mathcal U(t)\\
            \theta_i(t)&\mathrm{otherwise}
            \end{cases}\\
\end{cases}
\label{eq:discrete-evo}\end{equation} where the map
\(f_\varepsilon(z):\mathbb R\rightarrow\mathbb R\) with the sensitivity
parameter \(\varepsilon>0\) is defined as \[
f_\varepsilon(z)=\begin{cases}
z&\mathrm{if}\ |z|\geq\varepsilon\\
0&\mathrm{otherwise}.
\end{cases}
\] Also, let \(\mathrm{ave}_i(t)\) be the weighted average of the
relative states of the neighbors given by \begin{equation}
\mathrm{ave}_i(t) = \sum_{j\in\mathcal N_i}a_{ij}(t)(\widehat x_j(t-\tau_j^i(t)) -x_i(t))
\label{eq:aveit}\end{equation} where \(a_{ij}(t)\) is the \((i,j)\)th
entry of the (possibly time-varying) adjacency matrix
\(A(t)\in\mathbb R^{n\times n}\) corresponding to \(\mathcal G\), which
satisfies \(\alpha\leq a_{ij}(t)<1\) when \(a_{ij} \neq 0\) and
\(\alpha\) is the lower bound with \(0<\alpha\leq 1/2\). Furthermore,
\(\tau_j^i(t)\) denotes the time delay in the communication from agent
\(j\) to agent \(i\) at time \(t\). The variable \(\widehat x_j(t)\) is
the state value of agent \(j\) most recently received and stored by
agent \(i\). More specifically, it is given by \[
\widehat x_j(t) = x_j(t_k^j),\ \mathrm{for}\ t_k^j<t\leq t_{k+1}^j.
\] Additionally for \(t_m^i<t\leq t_{m+1}^i\), we define the length
\(e_i(t) = t - t_m^i\) of time interval since agent \(i\)'s last update.
Regarding the delay time in communication, we assume that it is upper
bounded for any \(i,j,t\) by \(\tau^\prime\) as \begin{equation}
e_i(t)+\tau_j^i(t)\leq\tau^\prime.
\label{eq:tauprime}\end{equation}

The algorithm for each agent \(i\) based on the evolutions
\cref{eq:continuous-evo,eq:discrete-evo} can be described as follows:
Agent \(i\) has the local clock \(\theta_i(t)\), which decreases as time
passes by. When its clock reaches 0 at time \(t\), it computes the
weighted average \(\mathrm{ave}_i(t)\) based on the states
\(\widehat{x}_j(t)\) received from its neighbors so far. Then, it sends
its updated state value \(x_i(t) = \widehat{x}_i(t)\) to its neighbors.

This protocol differs from the one proposed in {[}14{]} in not using
\(x_j(t)\) but using \(\widehat x_j(t)\) to update. This means that
there is no need for agents to send request to neighbors to receive the
neighbors' value at time \(t\). As a result, we can apply the protocol
to directed graphs.

\hypertarget{dynamics-in-the-discrete-system-formulation}{%
\subsubsection{Dynamics in the discrete system
formulation}\label{dynamics-in-the-discrete-system-formulation}}

In the dynamics of the hybrid protocol discussed above, for each agent
\(i\), the interval time length of the discrete evolution times is at
least \(\varepsilon\). Also the variable \(x(t)\) at time
\(t_l<t\leq t_{l+1}\) follows \(x(t_l) \leq x(t) \leq x(t_{l+1})\) or
\(x(t_l)\geq x(t) \geq x(t_{l+1})\). We can interpret the dynamics as a
discrete-time system by focusing on the evolution at times \(\{t_l\}\).

To simplify the notation, we denote the variables with discrete time
\(k\in\mathbb{Z}_+\) as \[
\begin{array}{ll}
x[k] = x(t_k),&u[k] = u(t_k)
\end{array}
\] and let the sequence \(\{k_m^i\}_{m\in\mathbb Z_+}\) be \[
\{k_m^i\}_{m\in\mathbb Z_+}=\{k:i\in\mathcal U[k]\},\ k_0^i = 0.
\]

Then the self-triggered algorithm for agent \(i\) can be represented in
discrete time as \begin{equation}
x_i[k+1] = \begin{cases}
\widehat x_i[k] + f_\varepsilon(\mathrm{dave}_i[k])&\mathrm{if}\ i\in{\mathcal U}[k]\\
x_i[k]+u_i[k](t_{k+1}-t_k)&\mathrm{otherwise}.
\end{cases}
\label{eq:tilde-evo}\end{equation} Further, the update rule of
\(\widehat x_i[k]\) is \begin{equation}
\widehat x_i[k+1] = \begin{cases}
\widehat x_i[k] + f_\varepsilon(\mathrm{dave}_i[k])&\mathrm{if}\ i\in{\mathcal U}[k]\\
\widehat x_i[k] &\mathrm{otherwise}
\end{cases}
\label{eq:hat-evo}\end{equation} where \(\mathrm{dave}_i[k]\) is defined
as \begin{equation}
\mathrm{dave}_i[k] = \sum_{j\in\mathcal N_i}a_{ij}[k-e_i[k]](\widehat x_j[k-e_i[k]-\tau_j^i[k]] -\widehat x_i[k])
\label{eq:avek}\end{equation} and regarding the delay time in
communication, we assume that it is upper bounded by \(\tau\) for the
given \(\tau^\prime\) in \cref{eq:tauprime} and any \(i,j,k\) as
\begin{equation}
e_i[k]+\tau_j^i[k]\leq\tau < \frac{n\tau^\prime}{\varepsilon}.
\label{eq:tau}\end{equation}

\hypertarget{adversarial-model-and-resilient-consensus}{%
\subsection{Adversarial model and resilient
consensus}\label{adversarial-model-and-resilient-consensus}}

Now, we consider the situation where some of the nodes in the network
are faulty or even adversarial. The objective here is to keep the
nonfaulty, regular nodes from being affected by such adversarial ones.
Here, we introduce the class of adversarial nodes considered in this
paper.

The nodes in \(\mathcal V\) are partitioned into two sets:
\(\mathcal R\) denotes the set of regular nodes and
\(\mathcal A = \mathcal V\mathbin{\backslash}\mathcal R\) represents the
set of adversarial nodes. The regular nodes will update their controls
\(u_i\) following the designed algorithm exactly while the adversarial
nodes can update \(u_i\) arbitrary. Here, let
\(n_R=|\mathcal R|,\ n_A=|\mathcal A|\). The attacker is allowed to know
the states of the regular nodes and the graph topology and to choose any
node as a member of \(\mathcal A\) under some constraints.

For the class of adversarial nodes, we employ the malicious model
defined as follows {[}6{]}.

\textbf{Definition 2.} (Malicious nodes): We say that an adversarial
node is malicious if it sends the same value to all of its neighbors at
each transmission.

Adversarial nodes more difficult to deal with are those that can send
different values to different neighbors in an arbitrary way. Such nodes
are called Byzantine nodes {[}7{]}.

In our problem setting, the identities of the malicious nodes are
unknown, but we assume the prior knowledge on the maximum number of
malicious nodes in the network. This is a common assumption employed in
the literature of fault-tolerant distributed algorithms {[}4{]}. More
specifically, we introduce the following models.

\textbf{Definition 3.} (\(F\)-total and \(F\)-local models): For
\(F\in\mathbb N\), the adversarial set \(\mathcal A\) follows the
\(F\)-total model if \(|\mathcal A|\leq F\), and the \(F\)-local model
if \(|\mathcal N_i \cap\mathcal A|\leq F\) for each node
\(i\in\mathcal R\).

This paper deals with networks of the \(F\)-total model.

In this context, the notion of resilient consensus for multi-agent
systems is now given {[}25{]}.

\textbf{Definition 4.} (Resilient consensus): Given \(c\geq 0\), the
multi-agent system is said to reach resilient consensus at the error
level \(c\) if for any possible sets and behaviors of the malicious
agents and any initial state values of the regular nodes, the following
conditions are satisfied.

\begin{enumerate}
\def\labelenumi{\arabic{enumi}.}
\tightlist
\item
  Safety condition: There exists an interval
  \(\mathcal S\subset\mathbb R\) determined by the initial states of the
  regular agents such that \(x_i[k]\in\mathcal S\) for all
  \(i\in\mathcal R, k\in\mathbb Z_+\).
\item
  Consensus condition: For all \(i,j\in\mathcal R\), it holds that
  \(\limsup_{k\rightarrow\infty}|x_i[k]-x_j[k]|\leq c\).
\end{enumerate}

The multi-agent problem of this paper can be formulated as follows:
Given \(c\geq 0\), design event-based update rules for the regular
agents so that they reach resilient consensus at the error level \(c\)
under the \(F\)-total model.

\hypertarget{self-triggered-consensus-protocol-1}{%
\section{SELF-TRIGGERED CONSENSUS
PROTOCOL}\label{self-triggered-consensus-protocol-1}}

\hypertarget{e-msr-algorithm}{%
\subsection{E-MSR algorithm}\label{e-msr-algorithm}}

As mentioned above, every agent updates its state value at every time by
the ternary input, but only when an update event happens, the auxiliary
values will be updated using values from its neighbors. For reaching
resilient consensus, we apply an algorithm where each regular agent
ignores some neighbors suspected to be behaving maliciously.

The algorithm given below follows the one from {[}25{]} called the
event-based mean subsequence reduced (E-MSR) algorithm.

\begin{enumerate}
\def\labelenumi{\arabic{enumi}.}
\tightlist
\item
  (Collecting neighbors' values) At time \(t\), every regular node
  \(i\in\mathcal R\) possesses the values \(\widehat x_i(t)\) (or
  \(\widetilde x_i(t)\)), most recently sent from the neighbors as well
  as its own value \(x_i(t)\) and sorts them in ascending order.
\item
  (Deleting suspicious values) Comparing with \(x_i(t)\), node \(i\)
  removes the \(F\) largest and \(F\) smallest values from the values
  collected at step 1. If the number of values larger or smaller than
  \(x_i(t)\) is less than \(F\), then all of them are removed. The
  removed data is considered as suspicious and will not be used in the
  update at the current time step. The set of node indices of the
  remaining values is written as \(\mathcal M_i(t)\subset\mathcal N_i\).
\item
  (Local update) Node \(i\) updates its auxiliary values with values
  sent from \(\mathcal M_i(t)\). The weight \(a_{ij}(t)\) of ignored
  neighbor \(j\) is set to 0 temporarily.
\end{enumerate}

\hypertarget{self-triggered-resilient-consensus-protocol}{%
\subsection{Self-triggered resilient consensus
protocol}\label{self-triggered-resilient-consensus-protocol}}

In the self-triggered consensus protocol based on the ternary
controller, to reach resilient consensus, the discrete evolution
\cref{eq:discrete-evo} can be replaced by introducing the E-MSR
algorithm with the following: \begin{equation}
\begin{cases}
  x_i(t^+) = x_i(t)\\
  u_i(t^+) = \begin{cases}
              \mathrm{sign}(f_\varepsilon(\mathrm{ave}_i^\mathrm{M}(t)))&\mathrm{if}\ i\in\mathcal U(t)\\
              u_i(t)&\mathrm{otherwise}
              \end{cases}\\
  \theta_i(t^+) = \begin{cases}
              \max\{|\mathrm{ave}_i^\mathrm{M}(t)|,\varepsilon\}&\mathrm{if}\ i\in\mathcal U(t)\\
              \theta_i(t)&\mathrm{otherwise}
              \end{cases}\\
  \end{cases}
\label{eq:discrete-evo-msr}\end{equation} \begin{equation}
\mathrm{ave}_i^\mathrm{M}(t) = \sum_{j\in\mathcal M_i(t)}a_{ij}(t)(\widehat x_j(t-\tau_j^i(t)) -x_i(t))
\label{eq:aveiMt}\end{equation}

Also, in the discrete-time system form, \cref{eq:tilde-evo} and
\cref{eq:hat-evo} can be replaced with \begin{equation}
x_i[k+1] = \begin{cases}
\widehat x_i[k] + f_\varepsilon(\mathrm{dave}_i^\mathrm{M}[k])&\mathrm{if}\ i\in{\mathcal U}[k]\\
x_i[k]+u_i[k](t_{k+1}-t_k)&\mathrm{otherwise}
\end{cases}
\label{eq:tilde-evo-msr}\end{equation} \begin{equation}
\widehat x_i[k+1] = \begin{cases}
\widehat x_i[k] + f_\varepsilon(\mathrm{dave}_i^\mathrm{M}[k])&\mathrm{if}\ i\in{\mathcal U}[k]\\
\widehat x_i[k]&\mathrm{otherwise}
\end{cases}
\label{eq:hat-evo-msr}\end{equation} \begin{equation}
\mathrm{dave}_i^\mathrm{M}[k] = \sum_{j\in{\mathcal M}_i[k]} a_{ij}[k-e_i[k]](\widehat x_j[k-e_i[k]-\tau_j^i[k]] -\widehat x_i[k]).
\label{eq:aveiMk}\end{equation}

For ease of notation, we reorder the indices of the agents. Let the
regular agents take indices \(1,...,n_R\) and let the malicious agents
be \(n_R+1,...,n\). Then the system variables are partitioned into the
regular and malicious parts as
\(x[k] = \left[x^R[k]^T\ x^A[k]^T\right]^T\).

As the neighbors' information can be delayed as shown in \cref{eq:tau},
each agent \(i\) updates \(x_i\) with the state values up to \(\tau\)
steps before. Thus let \[
\begin{array}{l}
\widehat z[k] = \left[\widehat x[k]^T\ \widehat x[k-1]^T\ \cdots\ \widehat x[k-\tau]^T\right]^T
\end{array}
\] with the initial values
\(\widehat{x}[-1] = \ldots = \widehat{x}[-\tau] = \widehat{x}[0]\). And
rewrite the update rule \cref{eq:hat-evo-msr} for \(\widehat x\) as
\begin{equation}
\widehat x_i[k+1] = \begin{cases}
  \left[W_\tau[k]\widehat z[k]\right]_i&\mathrm{if}\ i\in\mathcal R\ \mathrm{and}\\
  &\left|\left[W_\tau[k]\widehat z[k]\right]_i-\widehat x_i[k]\right|\geq\varepsilon\\
  \widehat x_i[k] & \mathrm{if}\ i\in\mathcal R\ \mathrm{and}\\
  &\left|\left[W_\tau[k]\widehat z[k]\right]_i-\widehat x_i[k]\right|<\varepsilon\\
  \widehat x_i[k] + u_i^A[k] & \mathrm{otherwise}.
\end{cases}
\label{eq:convex-form}\end{equation} Here, \(W_\tau[k]\) is the
\(n\times(\tau+1)n\) matrix defined as follows: \[
W_\tau[k] = \left[I_n\ 0\right]-L_\tau[k]
\] where \(L_\tau[k]\) is given by \[
L_\tau[k]=\left[\begin{array}{lllll}
D[k]-A_0[k]&-A_1[k]&\cdots&-A_\tau[k]
\end{array}\right].
\] with \[
\begin{aligned}
D[k]&=\mathrm{diag}(a_{1j}[k-e_1[k]],...,a_{nj}[k-e_n[k]]),\\
A_l[k]&=\begin{cases}
  a_{ij}[k-e_i[k]] & \mathrm{if}\ i\neq j\ \mathrm{and}\ e_i[k]+{\tau}_j^i[k]=l\\
  0&\mathrm{otherwise}.
\end{cases}
\end{aligned}
\] The input for the malicious agents is given as
\(u^A[k]\in\mathbb R^{n_A}\), which can be manipulated arbitrarily,
Here, let \(\omega\) be the minimum nonzero element of all possible
cases of \(W_\tau[k]\) over all \(k\). By the bound \(\alpha\) on
\(a_{ij}[k]\), we have \(\omega\in(0, 1)\).

Then we are ready to present the algorithm that each agent follows to
reach resilient consensus. The protocol given in Algorithm 1 represents
the self-triggered resilient consensus protocol based on the ternary
controller.

\begin{algorithm}[t]
\caption{$P_\mathrm{slf}^M$: Continuous Evolution for agent $i\in\mathcal R$}
\begin{algorithmic}[1]
  \State\textbf{Initialize:} {$u_i(0)\in\{-1,0,1\},\ \theta_i(0)=0$}
  \If{agent $i$ receives $x_j(t)$ from agent $j\in\mathcal N_i$}
    \State agent $i$ stores $x_j(t)$
  \EndIf
  \If{$\theta_i(t)=0$}
    \If{$u_i(t)\neq 0$}
      \ForAll{$j\in\mathcal N_i$}
        \State agent $i$ sends $x_i(t)$ to agent $j$
      \EndFor
    \EndIf
    \State \textbf{apply E-MSR setp 2} and set $\mathcal M_i(t)$
    \State set $\theta_i(t^+)=\max\{|\mathrm{ave}_i^\mathrm{M}(t)|,\varepsilon\}$
    \State set $u_i(t^+)=\mathrm{sign}(f_\varepsilon(\mathrm{ave}_i^\mathrm{M}(t)))$
  \EndIf
\end{algorithmic}
\end{algorithm}

\hypertarget{resilient-consensus-analysis}{%
\subsection{Resilient consensus
analysis}\label{resilient-consensus-analysis}}

By adopting the MSR technique in the ternary event-based control for
consensus as in \cref{eq:discrete-evo-msr}, we obtain the following main
result of this paper.

\textbf{Theorem 1}: Under the \(F\)-total malicious model, the regular
agents with the E-MSR algorithm in the multi-agent system reach
resilient consensus at error level \(c\) if the network topology is
\((2F+1)\)-robust. The safety interval is given by
\(\mathcal S=[\min x^R[0], \max x^R[0]]\), and the error level \(c\) is
achieved if the parameter \(\varepsilon\) satisfies \[
\varepsilon\leq\frac{\omega^{(\tau+1)n-1}(1-\omega)c}{1-\omega^{(\tau+1)n-1}}.
\]

\textbf{Proof}: We show that the update rule in \cref{eq:tilde-evo-msr}
meets both safety and consensus conditions.

\hypertarget{safety-condition}{%
\subsubsection{Safety condition}\label{safety-condition}}

We must prove that the state value of regular agents satisfy
\(x_i[k]\in\mathcal S\) for all \(i\in\mathcal R, k\in\mathbb Z_+\).
First, we define \[
\overline{\widehat z}[k] = \max\widehat z^R[k],\ \underline{\widehat z}[k] = \min\widehat z^R[k].
\] By \cref{eq:convex-form}, \(\widehat x_i[k+1]\) is given by the
convex combination of the entries of \(\widehat z[k]\) for the regular
agent \(i\) at step \(k\). If some of the neighbors are malicious and
lie outside the range of the regular agents' values, then they will be
ignored by step 2 in the E-MSR algorithm. Consequently,
\(\max \widehat x^R[k+1]\leq\overline{\widehat z}[k]\) always holds.
Hence, \(\overline{\widehat z}[k+1]\) meets \[
\begin{aligned}
\overline{\widehat z}[k+1]&=\max(\widehat x^R[k+1],\widehat x^R[k],\ldots,\widehat x^R[k+1-\tau])\\
&\leq\max(\widehat x^R[k+1],\widehat x^R[k],\ldots,\widehat x^R[k-\tau])\\
&\leq\max(\widehat x^R[k],\ldots,\widehat x^R[k-\tau])\ =\ \overline{\widehat z}[k].
\end{aligned}
\] As a result, \(\overline{\widehat z}[k]\) is nonincreasing and it
holds that
\(\widehat x_i[k]\leq\overline{\widehat z}[k]\leq\overline{\widehat z}[0]\).
Further, at each step \(k\in(k_l^i,k_{l+1}^i]\) for the regular agent
\(i\) by the event-based control, it holds either
\(x_i[k_l^i] \leq x_i[k]\leq x_i[k_{l+1}^i]\) or
\(x_i[k_{l+1}^i] \leq x_i[k]\leq x_i[k_l^i]\). Thus, it holds:
\begin{equation}
\begin{aligned}
x_i[k]&\leq\max(\widehat z^R[k_l^i],\widehat z^R[k_{l+1}^i])\\
&\leq\overline{\widehat z}[k_l^i]\leq\overline{\widehat z}[0]=\max x^R[0].
\end{aligned}
\label{eq:xleqz0}\end{equation}

Likewise, we can show \begin{equation}
\begin{array}{ll}
\widehat x_i[k]\geq\min x^R[0],&x_i[k]\geq\min x^R[0]
\end{array}
\label{eq:xgeqz0}\end{equation} and thus the safety condition is
satisfied.

\hypertarget{consensus-condition}{%
\subsubsection{Consensus condition}\label{consensus-condition}}

We first sort the regular agents' values in the vector
\(\widehat z^R[k]\) at step \(k\) in ascending order. Denote by
\(s_i[k]\) the index of the agent taking the \(i\)th value from the
smallest. Hence, the values are sorted as
\(\widehat z_{s_1}[k]\leq\widehat z_{s_2}[k]\leq\cdots\leq\widehat z_{s_{(\tau+1)n}}[k]\).

Introduce two sequences of conditions for the bound of gaps between two
agents. The condition sequences \(\{\mathrm{ASC}_l\}\) and
\(\{\mathrm{DESC}_l\}\) are defined as follows:

\begin{itemize}
\tightlist
\item
  \(\mathrm{ASC}_1: \widehat z_{s_2}[k]-\widehat z_{s_1}[k]\leq\varepsilon/\omega\)
\item
  \(\mathrm{ASC}_2: \widehat z_{s_3}[k]-\widehat z_{s_2}[k]\leq\varepsilon/\omega^2\)
\item
  \ldots{}
\item
  \(\mathrm{ASC}_{(\tau+1)n-1}: \widehat z_{s_{(\tau+1)n}}[k]-\widehat z_{s_{(\tau+1)n-1}}[k]\leq\varepsilon/\omega^{(\tau+1)n-1}\)
\end{itemize}

\begin{itemize}
\tightlist
\item
  \(\mathrm{DESC}_1: \widehat z_{s_{(\tau+1)n}}[k]-\widehat z_{s_{(\tau+1)n-1}}[k]\leq\varepsilon/\omega\)
\item
  \(\mathrm{DESC}_2: \widehat z_{s_{(\tau+1)n-1}}[k]-\widehat z_{s_{(\tau+1)n-2}}[k]\leq\varepsilon/\omega^2\)
\item
  \ldots{}
\item
  \(\mathrm{DESC}_{(\tau+1)n-1}: \widehat z_{s_2}[k]-\widehat z_{s_1}[k]\leq\varepsilon/\omega^{(\tau+1)n-1}\).
\end{itemize}

\noindent Now, denote by \(j_A\) the minimum \(j,\ 1\leq j< (\tau+1)n\),
such that the condition \(\mathrm{ASC}_j\) is not satisfied. Also,
denote by \(j_D\) the maximum \(j,\ 1<j\leq (\tau+1)n\), such that the
condition \(\mathrm{DESC}_j\) is not satisfied. Then we have
\begin{equation}
\begin{aligned}
\widehat z_{s_{j_A+1}}[k]-\widehat z_{s_{j_A}}[k]&>\frac{\varepsilon}{\omega^{j_A}}\\
\widehat z_{s_{j_D}}[k]-\widehat z_{s_{j_D-1}}[k]&>\frac{\varepsilon}{\omega^{(\tau+1)n-j_D+1}}.
\end{aligned}
\label{eq:ajfalse}\end{equation} Furthermore, the conditions
\(\mathrm{ASC}_1\) to \(\mathrm{ASC}_{j_A-1}\) and
\(\mathrm{DESC}_{j_D+1}\) to \(\mathrm{DESC}_{(\tau+1)n-1}\) are
satisfied. Also, for \(0\leq k\leq k^\prime\), we introduce the
following sets: \begin{equation}
\begin{array}{l}
\mathcal X_1(k,k^\prime) = \left\{ j\in\mathcal R:\widehat x_j[k^\prime]<\widehat z_{s_{j_A}}[k]+\varepsilon \right\}\\
\mathcal X_2(k,k^\prime) = \left\{ j\in\mathcal R:\widehat x_j[k^\prime]>\widehat z_{s_{j_D}}[k]-\varepsilon \right\}.
\end{array}
\label{eq:mathcalx12}\end{equation}

By the nature of these two sets, every entry of the condition sequences
\(\mathrm{ASC}, \mathrm{DESC}\) is finally satisfied as step grows and
we can show the consensus condition. We study them separately according
to the relation of \(j_A, j_D\) and
\(\mathcal X_1(k, k),\mathcal X_2(k, k)\).

\textbf{Case 1. \(j_A<j_D\)}: There are four subcases, denoted by (1-a)
to (1-d), which are separately studied below.

\textbf{(1-a).
\(\mathcal X_1(k,k)\neq\phi, \mathcal X_2(k,k)\neq\phi\)}: For a regular
agent \(j\notin\mathcal X_1(k, k)\), by definition, it holds \[
\widehat x_j[k]\geq\widehat z_{s_{j_A}}[k]+\varepsilon
\] and \(\widehat x_j[k]\geq\widehat z_{s_{j_A+1}}[k]\) since the
minimum element of \(\widehat z\) that exceeds
\(\widehat z_{s_{j_A}}[k]+\varepsilon\) is
\(\widehat z_{s_{j_A+1}}[k]\). If \(j\in{\mathcal U}[k]\), then values
less than \(\underline{\widehat z}[k] = \widehat z_{s_1}[k]\) will be
ignored. Additionally since the update is based on the convex
combination as shown in \cref{eq:convex-form} and \(\omega\) is the
lower bound of the elements of \(W_\tau[k]\), it holds \begin{equation}
\widehat x_j[k+1]\geq (1-\omega)\widehat z_{s_1}[k]+\omega\widehat z_{s_{j_A+1}}[k].
\label{eq:betaconvex}\end{equation} Using the conditions
\(\mathrm{ASC}_1\) to \(\mathrm{ASC}_{j_A-1}\), we can bound
\(\widehat{z}_{s_1}\) from below as \[
\begin{aligned}
\widehat z_{s_1}[k]&\geq\widehat z_{s_2}[k]-\frac{\varepsilon}{\omega}\geq\widehat z_{s_3}[k]-\left(\frac{1}{\omega}+\frac{1}{\omega^2}\right)\varepsilon\\
&\geq\cdots\geq\widehat z_{s_{j_A}}[k]-\left(\frac{1}{\omega}+\frac{1}{\omega^2}+\cdots+\frac{1}{\omega^{j_A-1}}\right)\varepsilon.
\end{aligned}
\] Substituting this into \cref{eq:betaconvex}, we can obtain
\begin{equation}
\begin{aligned}
\widehat x_j[k+1]&\geq\widehat z_{s_{j_A}}[k]+\omega(\widehat z_{s_{j_A+1}}[k]-\widehat z_{s_{j_A}}[k])-\frac{\varepsilon}{\omega^{j_A-1}}+\varepsilon\\
&>\widehat z_{s_{j_A}}[k]+\frac{\omega\varepsilon}{\omega^{j_A}}-\frac{\varepsilon}{\omega^{j_A-1}}+\varepsilon=\widehat z_{s_{j_A}}[k]+\varepsilon.
\end{aligned}
\label{eq:jgooutfromx1}\end{equation}

On the other hand, for an agent \(j\notin{\mathcal U}[k]\), we have \[
\widehat x_j[k+1] = \widehat x_j[k]>\widehat z_{s_{j_A}}[k]+\varepsilon.
\] Thus, we can derive \begin{equation}
j\notin\mathcal X_1(k,k)\Rightarrow j\notin\mathcal X_1(k,k+1).
\label{eq:jnotinx1}\end{equation} This means that if an agent does not
belong to \(\mathcal X_1(k,k)\), then it will not belong to
\(\mathcal X_1(k,k^\prime)\) at step \(k^\prime>k\) either. Likewise,
for \(\mathcal X_2(k,k)\), we can show \begin{equation}
j\notin\mathcal X_2(k,k)\Rightarrow j\notin\mathcal X_2(k,k+1).
\label{eq:jnotinx2}\end{equation}

Next, \(\mathcal X_1(k,k)\) and \(\mathcal X_2(k,k)\) are disjoint
because \[
\widehat z_{j_D}[k]-\widehat z_{j_A}[k]>\max\left\{\frac{1}{\omega^{j_A}},\frac{1}{\omega^{(\tau+1)n-j_D+1}}\right\}\varepsilon\geq2\varepsilon.
\] Thus, the graph being \((2F+1)\)-robust guarantees that there is at
least one agent \(i\) satisfying \(i\in\mathcal X_1(k,k)\) with
\(|\mathcal N_i\mathbin{\backslash}\mathcal X_1(k,k)|\geq 2F+1\) or
\(i\in\mathcal X_2(k,k)\) with
\(|\mathcal N_i\mathbin{\backslash}\mathcal X_2(k,k)|\geq 2F+1\). Here
we first consider the case where \(i\in\mathcal X_1(k,k)\). There exists
\(l\) such that \(k\leq k_l^i\leq k+\tau\) and
\(\widehat x_i[k^\prime+1] = \widehat x_i[k^\prime]\) at
\(k\leq k^\prime< k_l^i\). And thus for agent \(i\), it holds that
\(i\in\mathcal X_1(k,k_l^i)\). By \cref{eq:jnotinx1}, it also holds that
\(|\mathcal N_i\mathbin{\backslash}\mathcal X_1(k,k_l^i)|\geq 2F+1\).
Hence, in the update of agent \(i\) at step \(k_l^i\), there is at least
one neighbor \(j\) whose value satisfies
\(\widehat x_j[k_l^i]>\widehat z_{s_{j_A}}[k]+\varepsilon\). Thus like
\cref{eq:jgooutfromx1}, we have \begin{equation}
\widehat x_i[k_l^i+1]\geq (1-\omega)\widehat z_{s_1}[k]+\omega\widehat z_{s_{j_A+1}}[k]>\widehat z_{s_{j_A}}[k]+\varepsilon
\label{eq:iinx1}\end{equation} where it holds
\(\left|\left[W_\tau[k_l^i]\widehat z[k_l^i]\right]_i-\widehat x_i[k_l^i]\right|\geq\varepsilon\)
because \(\widehat x_i[k_l^i]\leq\widehat z_{s_{j_A}}[k]\). In this way,
using \cref{eq:jnotinx1} and \cref{eq:iinx1}, we have \begin{equation}
|\mathcal X_1(k,k+\tau)|\leq|\mathcal X_1(k,k_l^i)|<|\mathcal X_1(k,k)|
\label{eq:x1decreases}\end{equation} and thus after \(\tau\) steps, the
cardinality of \(\mathcal X_1(k,k+\tau)\) is smaller than that of
\(\mathcal X_1(k,k)\).

Similar results also hold for the case \(i\in\mathcal X_2(k,k)\), and
the cardinality of \(\mathcal X_2(k,k+\tau)\) will be smaller than that
of \(\mathcal X_2(k,k)\) after \(\tau\) steps. If the two sets are
nonempty, we can repeat the steps above. As a result,
\(|\mathcal X_1(k,k+\tau n)|=0\) or \(|\mathcal X_2(k,k+\tau n)|=0\) is
derived. It means that \(\overline{\widehat z}\) will decrease by
\(\varepsilon\) (or \(\underline{\widehat z}\) will increase by
\(\varepsilon\)) after \(\tau n\) steps.

\textbf{(1-b). \(\mathcal X_1(k,k)=\phi, \mathcal X_2(k,k)\neq\phi\)}:
It holds that \(\mathcal X_1(k,k+\tau)=\phi\) because of
\cref{eq:jnotinx1}. Therefore \[
\begin{aligned}
\underline{\widehat z}[k+\tau]&=\min(\widehat x^R[k],\widehat x^R[k+1],\ldots,\widehat x^R[k+\tau])\\
&\geq\widehat z_{s_{j_A}}[k]+\varepsilon\geq\widehat z_{s_1}[k]+\varepsilon
\end{aligned}
\] and \(\underline{\widehat z}\) increases by \(\varepsilon\) as
\(\tau\) steps goes by.

\textbf{(1-c). \(\mathcal X_1(k,k)\neq\phi, \mathcal X_2(k,k)=\phi\)}:
It holds that \(\underline{\widehat z}\) increases by \(\varepsilon\)
after \(\tau\) steps similarly to Case (1-b).

\textbf{(1-d). \(\mathcal X_1(k,k)=\phi, \mathcal X_2(k,k)=\phi\)}: In
this last case also, \(\underline{\widehat z}\) increases by
\(\varepsilon\) and \(\underline{\widehat z}\) increases by
\(\varepsilon\) as \(\tau\) steps go by.

\textbf{Case 2. \(j_A\geq j_D\)}: This case is in fact impossible. This
is because it would imply that \(\mathrm{ASC}_{j_D-1}\) and
\(\mathrm{DESC}_{j_A-1}\) are both satisfied. Thus we have
\(\varepsilon/\omega^{(\tau+1)n-j_D+1}<\widehat x_{s_{j_D}}[k]-\widehat x_{s_{j_D-1}}[k]\leq\varepsilon/\omega^{j_D-1}\)
and
\(\varepsilon/\omega^{j_A}<\widehat x_{s_{j_A+1}}[k]-\widehat x_{s_{j_A}}[k]\leq\varepsilon/\omega^{(\tau+1)n-j_A}\).
These inequalities indicate that it must hold \(j_D>(n+1)/2\) and
\(j_A<n/2\). Consequently, we have \(j_A< j_D\) and it contradicts with
the condition \(j_A\geq j_D\).

If \(j_A\) and \(j_D\) continuously exist, then
\(\overline{\widehat z}[k] - \underline{\widehat z}[k]\) will become
smaller and eventually negative, which cannot happen. Hence, we can
conclude that after finite time steps, all conditions of
\(\{\mathrm{ASC}_l\}\) and \(\{\mathrm{DESC}_l\}\) are satisfied to stop
the value \(\overline{\widehat z}[k]\) decreasing and
\(\underline{\widehat z}[k]\) increasing. This completes the
proof.$\qed$

The proof technique used above is inspired by the one in {[}25{]}, which
is for event-triggered communication in resilient consensus. There, we
have provided two approaches for the consensus algorithms. In this
paper, we adopted one of them (Protocol 2), for which a more tight
result can be obtained for approximate consensus. The result in Theorem
1 is more general, allowing time delays in the communications, though
with no delay we can establish a necessary and sufficient condition for
resilient consensus, stated in terms of \((F+1,F+1)\)-robustness,
instead of the sufficient condition in Theorem 1. From the technical
viewpoint, the presence of delays necessitates a change in the sequences
of conditions \(\mathrm{ASC}\) and \(\mathrm{DESC}\). In particular,
they are stated in terms of the value \(\widehat{z}\), which includes
delayed values. As a result, we must consider the four subcases in Case
1 in the proof. When no delay is present, these subcases reduces to one
since neither \(\mathcal X_1(k, k)\) nor \(\mathcal X_2(k, k)\) may be
empty.

It is also interesting to note that the framework in {[}25{]} can be
extended to ternary controller of {[}14{]} which operates in continuous
time. This is because the consensus algorithm under ternary controller
can be expressed as a hybrid system, which can be further interpreted as
a discrete-time system by focusing on the times when events take place.
This proof approach is different from that in {[}14{]}, which also
considers networks operating on undirected graphs.

\hypertarget{event-triggered-consensus-protocol}{%
\section{EVENT-TRIGGERED CONSENSUS
PROTOCOL}\label{event-triggered-consensus-protocol}}

We provide the event-triggered resilient consensus protocol based on
ternary controller. This protocol is introduced to compare with the
self-triggered protocol because it works in the same setting as the
self-triggered one. In this protocol, agent \(i\) does not have the
local clock variable \(\theta_i\) but instead the triggering threshold
\(\eta_i\). Each agent triggers update events when it receives a value
and transmits events when its value \(x_i\) exceeds \(\eta_i\). Let
\(\mathcal U(t)\) be the set of agents which trigger a communication
event at time \(t\).

For the continuous time \(t\), the system
\((x, u, \eta)\in\mathbb R^{2n}\) satisfies the following continuous
evolution \begin{equation}
\begin{cases}
\dot x_i = u_i\\
\dot u_i = 0\\
\dot \eta_i = 0
\end{cases}
\label{eq:eve-continuous-evo}\end{equation} except for every \(t\) such
that the set \(\mathcal U(t)\) is non-empty.

To define the rule of triggering communication events, we define
\(\widetilde x_j(t)\) with \(t_k^j<t\leq t_{k+1}^j\) for agent \(j\) by
\[
\widetilde x_j(t) = x_j(t_k^j)
\] and let the value \(h_i(t)\) for agent \(i\) at time \(t\) be \[
h_i(t) = |\widetilde x_i(t) - x_i(t)| - \eta_i(t).
\] Here, agent \(i\) communicates at the time instants when \(h_i(t)\)
becomes \(0\) or more. Then at time instants \(t=t_k^j+\tau_j^i(t_k^j)\)
(\(k\geq0, j\in\mathcal N_i\)) for agent \(i\), the system satisfies the
following discrete evolution: \begin{equation}
\begin{cases}
x_i(t^+) = x_i(t)\\
u_i(t^+) = \mathrm{sign}(f_\varepsilon(\widetilde{\mathrm{ave}}_i^M(t)))\\
\eta_i(t^+) = \max\{|\widetilde{\mathrm{ave}}_i^\mathrm{M}(t)|,\varepsilon\}.
\end{cases}
\label{eq:eve-discrete-evo}\end{equation} where
\(\widetilde{\mathrm{ave}}_i^M(t)\) is given by \begin{equation}
\widetilde{\mathrm{ave}}_i^M(t) = \sum_{j\in\mathcal M_i(t)}a_{ij}(t)(\widetilde x_j(t-\tau_j^i(t)) -\widetilde x_i(t)).
\label{eq:tildeaveit}\end{equation}

The event-triggered resilient consensus protocol of these evolutions
\cref{eq:eve-continuous-evo} and \cref{eq:eve-discrete-evo} is the
following: Each agent \(i\) has own threshold \(\eta_i\). When its error
\(|\widetilde x_i - x_i|\) reaches \(\eta_i\), it sends the value
\(x_i=\widehat x_i\) to neighbors. At the time when an agent receives
values, it updates control \(u_i\) and \(\eta_i\) with value
\(x_j = \widetilde x_j\) to which recently send from its neighbors. This
is formally shown in Algorithm 2.

In the event-triggered consensus protocol based on ternary controller,
we can also prove to reach the same resilient consensus such as
self-triggered one.

\begin{algorithm}[t]
\caption{$P_\mathrm{eve}^M$: Continuous Evolution for agent $i\in\mathcal R$}
\begin{algorithmic}[1]
  \State\textbf{Initialize:} {$u_i(0)\in\{-1,0,1\},\ \eta_i(0) = 0$}
  \If{agent $i$ receives $x_j(t)$ from agent $j\in\mathcal N_i$}
    \State agent $i$ stores $x_j(t)$
    \State \textbf{apply E-MSR step 2} and set $\mathcal M_i(t)$
    \State set $u_i(t^+)=\mathrm{sign}(f_\varepsilon(\widetilde{\mathrm{ave}}_i^M(t)))$
    \State set $\eta_i(t^+) = \max\{|\widetilde{\mathrm{ave}}_i^\mathrm{M}(t)|,\varepsilon\}$
  \EndIf
  \If{$|\widetilde x_i(t) - x_i(t)| - \eta_i(t)\geq 0$}
    \State $\widetilde x_i(t) = x_i(t)$
    \ForAll{$j\in\mathcal N_i$}
      \State agent $i$ send $x_i(t)$ to agent $j$
    \EndFor
  \EndIf
\end{algorithmic}
\smallskip
\end{algorithm}

\hypertarget{numerical-example}{%
\section{NUMERICAL EXAMPLE}\label{numerical-example}}

In this section, we illustrate the effectiveness of the two proposed
resilient consensus protocols and compare their performance with the
conventional nonresilient counterpart.

\hypertarget{small-network-with-3-robust-property}{%
\subsection{Small network with 3-robust
property}\label{small-network-with-3-robust-property}}

\begin{figure}[t]
\hypertarget{fig:robust-example}{%
\centering
\includegraphics[width=\textwidth,height=22mm]{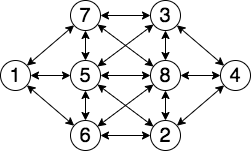}
\caption{3-robust graph}\label{fig:robust-example}
}
\end{figure}

First we consider the multi-agent system with 8 nodes whose connectivity
graph is shown in \cref{fig:robust-example}. We can check that it meets
the condition to be 3-robust. For the three protocols, we use common
initial states
\(x(0)=[0,\frac{1}{6},\frac{1}{3},\frac{1}{2},\frac{2}{3},\frac{5}{6},1,\frac{1}{2}]\).
The bound on time delay is set as \(\tau=0.1\). The adversary is taken
to be agent 8, which will continuously oscillate its state by following
a sine curve. We also took the sensitivity parameter
\(\varepsilon = 0.1\). Here, the malicious agent sends its values once
every \(\varepsilon\) time length.

\begin{figure}[t]
\hypertarget{fig:nomsr_01_01_comm}{%
\centering
\includegraphics{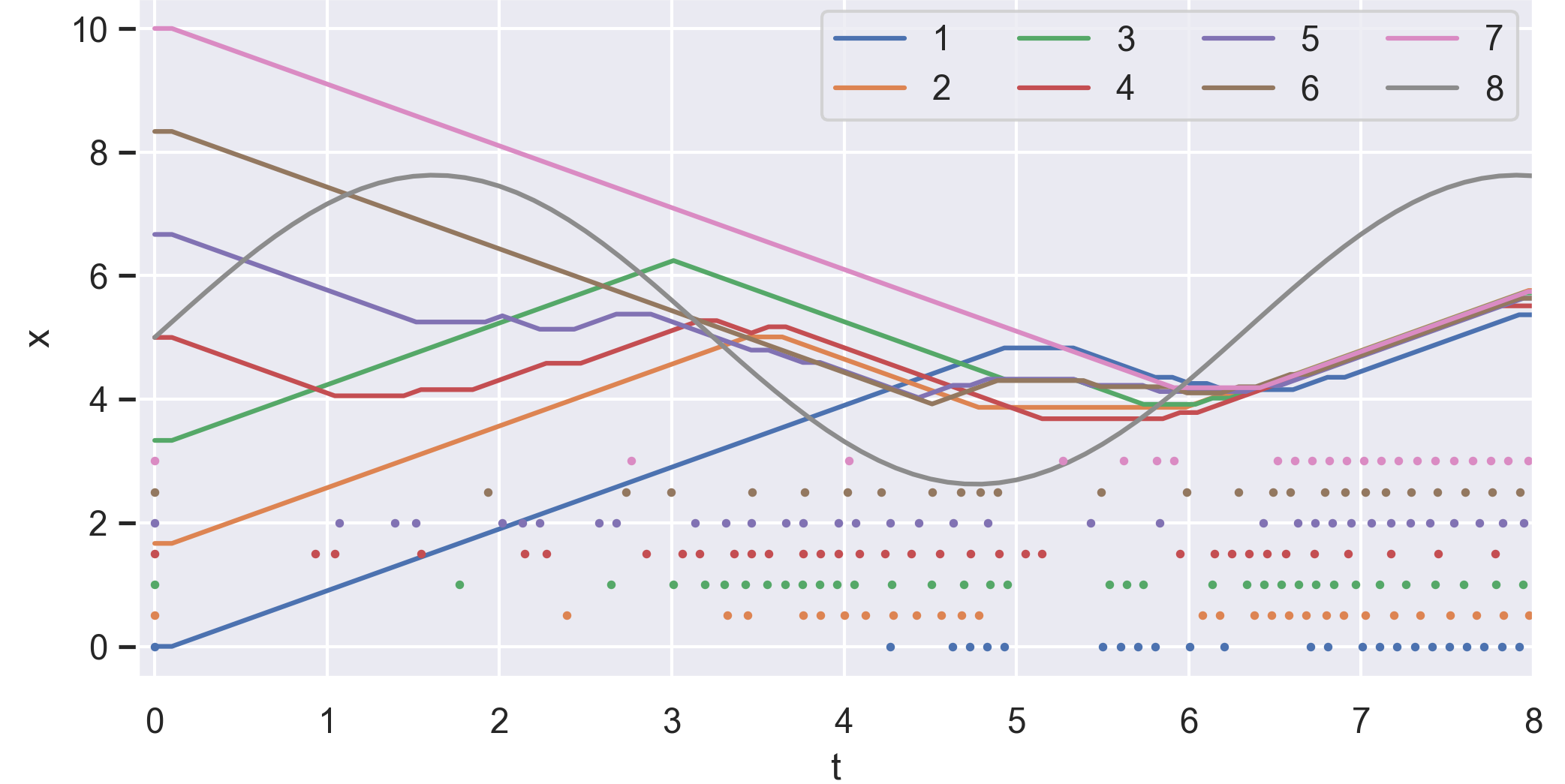}
\caption{Time responses for the conventional self-triggered
protocol}\label{fig:nomsr_01_01_comm}
}
\end{figure}

The conventional self-triggered protocol of {[}14{]} is not resilient
against the adversary. The time responses of the agents' states are
shown in \cref{fig:nomsr_01_01_comm}. Also, in the plot, the dots
indicate the transmission time instants of the agents, whose colors
match those of the states. It is clear that the agents are influenced by
the adversary. Their states change over time and transmissions do not
stop.

The simulation results for the proposed self-triggered and
event-triggered resilient protocols are shown in Figs. 4 and 5,
respectively. We observe that these protocols manages to achieve the
desired level of error in consensus. Moreover, the regular agents stop
transmitting after they reach consensus. These protocols are thus more
efficient in the amount of communication in comparison with the
sampling-based time-triggered protocols.

\begin{figure}[t]
\hypertarget{fig:self_01_01_comm}{%
\centering
\includegraphics{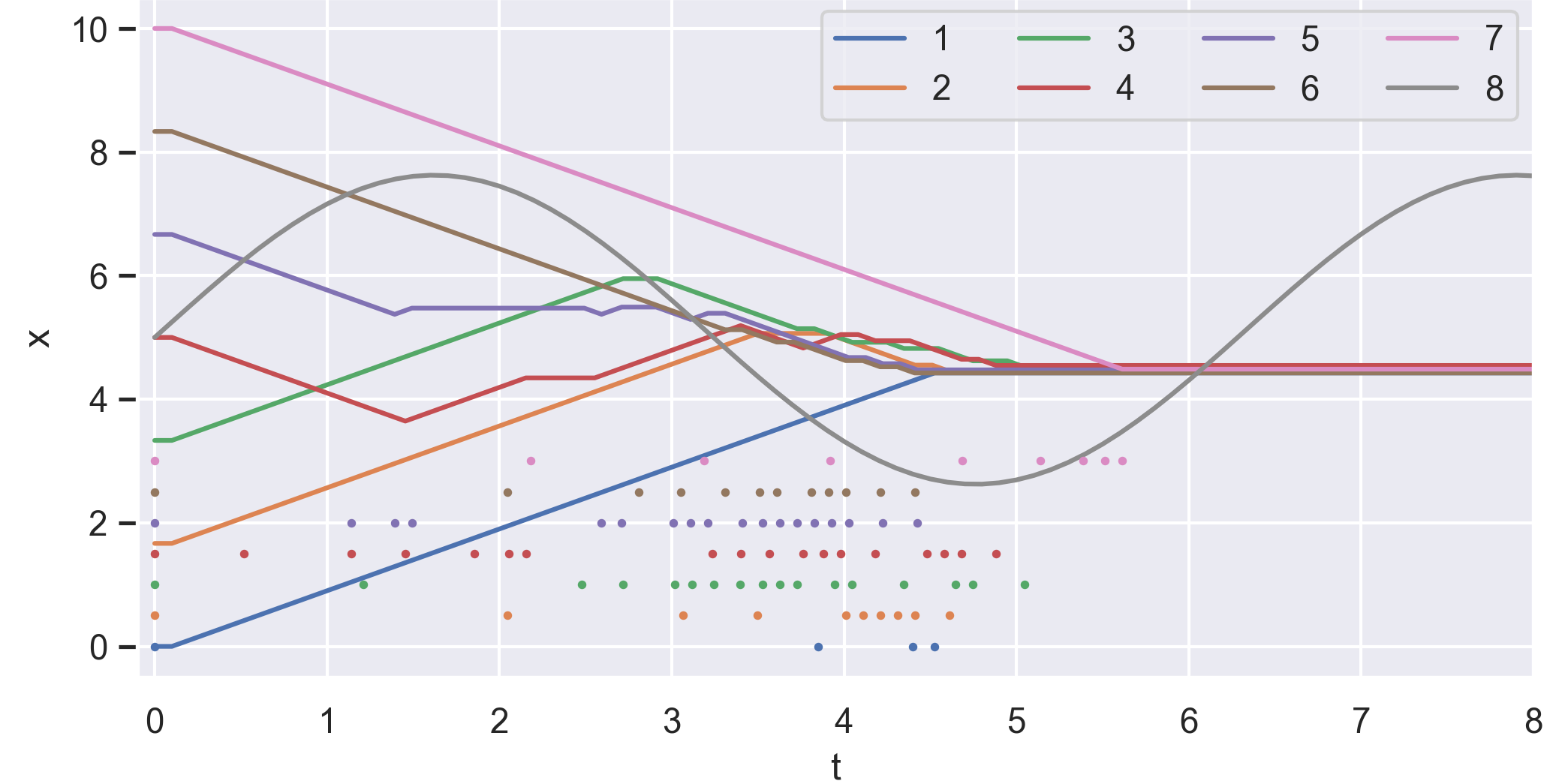}
\caption{Time responses for the self-triggered resilient
protocol}\label{fig:self_01_01_comm}
}
\end{figure}

\begin{figure}[t]
\hypertarget{fig:eve_01_01_comm}{%
\centering
\includegraphics{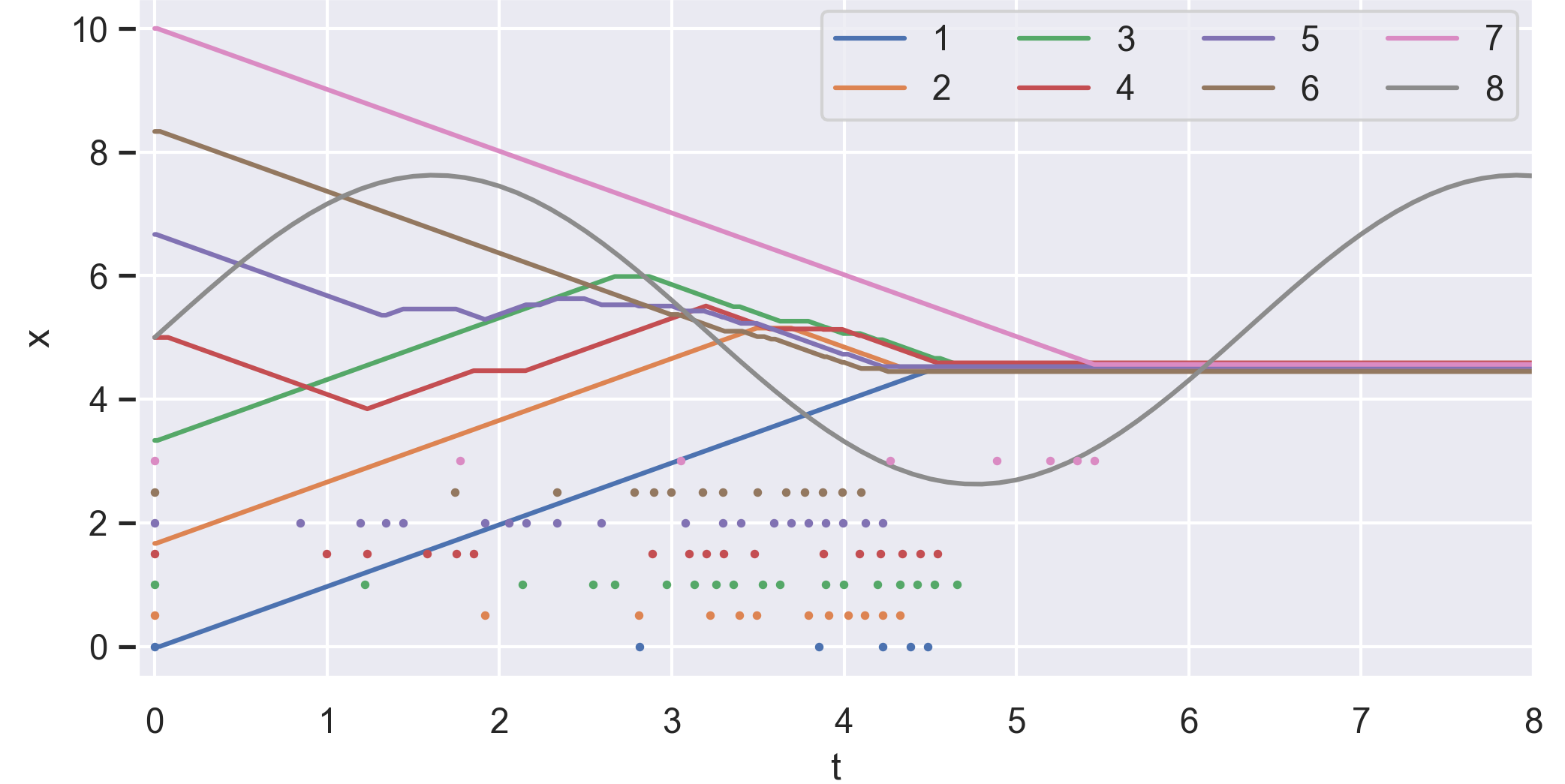}
\caption{Time responses for the event-triggered resilient
protocol}\label{fig:eve_01_01_comm}
}
\end{figure}

\hypertarget{random-networks}{%
\subsection{Random networks}\label{random-networks}}

Next, we consider a larger multi-agent system with 100 nodes whose
underlying graph is a random geometric network. We placed 100 nodes
randomly within a field of a unit square. The nodes have a common
communication range \(r\) and communicate with neighbors within the
range.

\hypertarget{consensus-success-rates}{%
\subsubsection{Consensus success rates}\label{consensus-success-rates}}

First, we demonstrate how the proposed protocols perform over such
random graphs where the robustness property cannot be guaranteed since
the network is too large. Here, we examined the communication range
\(r \in (0,0.5]\) and used different numbers for the malicious agents as
\(n_A =0,3,5\). The malicious agents here changed their controls
randomly within the range \([-10, 10]\) and sent a message every
\(\varepsilon\) time length, where the sensitivity parameter was set as
\(\varepsilon = 1\).

\begin{figure}[t]
\hypertarget{fig:prob_plot}{%
\centering
\includegraphics{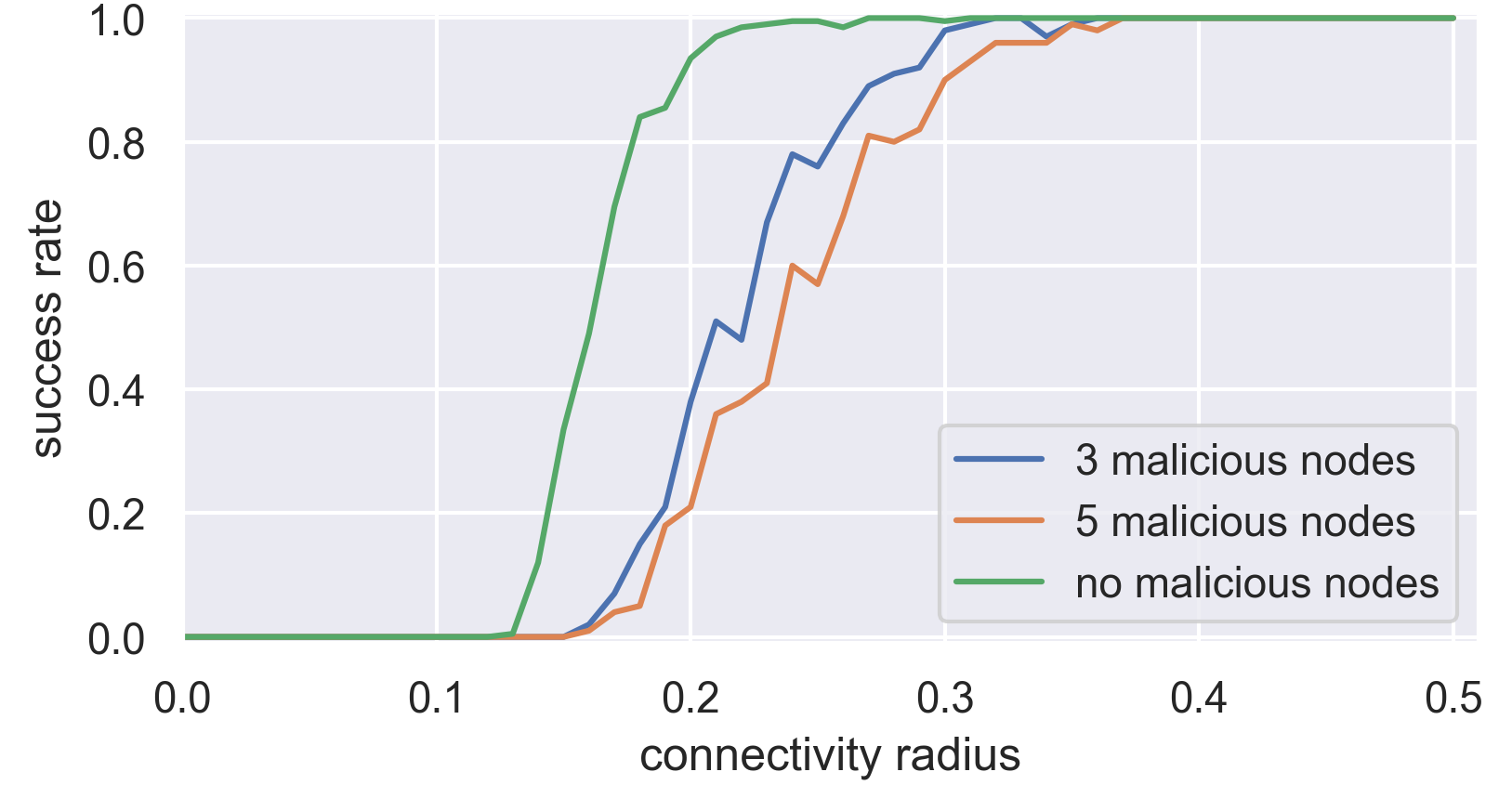}
\caption{Consensus success rates over 100 Monte Carlo simulations for
the self-triggered resilient protocol}\label{fig:prob_plot}
}
\end{figure}

\cref{fig:prob_plot} displays the success rates of consensus for the
three cases with \(n_A=0,3,5\) for the self-triggered resilient protocol
obtained from 100 Monte Carlo simulations. The success rate with no
malicious agent (\(n_A=0\)) matches the rate of the network being
connected. Observe that to achieve consensus with more malicious agents,
more connectivity is required for the network. However, it is clear that
the network need not be complete for the success rate to reach 1. The
results for the event-triggered control are almost the same and hence
not indicated in the plot.

\hypertarget{comparison-between-the-two-protocols}{%
\subsubsection{Comparison between the two
protocols}\label{comparison-between-the-two-protocols}}

\begin{table}[t]
\hypertarget{tbl:trig_and_comm}{%
\begin{center}
\caption{Average numbers of updates and transmissions for resilient protocols over 50 Monte Carlo simulations with $r=0.4$ and $\varepsilon=1$}
\begin{tabular}{|c|c|c|c|c|}
\hline
 & \multicolumn{2}{c|}{Self-triggered protocol} & \multicolumn{2}{c|}{Self-triggered protocol} \\ \hline
$n_A$ & \# updates & \# transmissions & \# updates & \# transmissions \\ \hline
0 & 20.3 & 90.8 & 91.0 & 88.4 \\ \hline
3 & 20.3 & 90.6 & 109 & 87.8 \\ \hline
5 & 20.3 & 89.2 & 121 & 88.6 \\ \hline
\end{tabular}
\end{center}
\label{tbl:trig_and_comm}
}
\end{table}

Next, we highlight the difference between the two resilient protocols,
which has not been evident in the theoretical results or from the
experiments so far. We show that the self-triggered approach is
advantageous in keeping the number of update events minimal. To this
end, we run simulations for both protocols and compare the numbers of
updates and transmissions per regular agent until time \(t=20\). Here,
we used the random network with the communication radius at \(r=0.4\).
Table I displays the average numbers over 50 Monte Carlo simulations.
The total numbers of messages sent by the regular agents were about the
same for both protocols and remain constant regardless of the size of
\(n_A\).

On the other hand, a clear difference arises in the numbers of updates.
For the event-triggered protocol, this number is in general larger than
that for the self-triggering protocol. However, we see that it is more
vulnerable to the malicious agents; the number of updates increases for
larger \(n_A\) since the agents must make an update each time a message
is received from neighbors and in particular, under the setting here,
the malicious agents transmitted messages very frequently. In contrast,
for the self-triggered protocol, the number did not change by \(n_A\)
and is one fifth of that for the other protocol.

\hypertarget{conclusion}{%
\section{CONCLUSION}\label{conclusion}}

We have proposed consensus protocols resilient to attacks from malicious
agents based on self- and event-triggered consensus protocols using
ternary control. The resilience is achieved by applying the MSR-type
techniques while the communication among the agents is efficient and
robust to networks in uncertain environments with limited communication
resources.

\parindent = 0pt

\footnotesize

\hypertarget{bibliography}{%
\section*{REFERENCES}\label{bibliography}}
\addcontentsline{toc}{section}{REFERENCES}

\hypertarget{refs}{}
\begin{cslreferences}
\leavevmode\hypertarget{ref-GraphTheory}{}%
{[}1{]} M. Mesbahi and M. Egerstedt, \emph{Graph Theoretic Methods in
Multiagent Networks}. Princeton University Press, 2010.

\leavevmode\hypertarget{ref-ResearchChallenges}{}%
{[}2{]} A. A. Cárdenas, S. Amin, and S. Sastry, ``Research challenges
for the security of control systems.'' in \emph{HotSec}, 2008.

\leavevmode\hypertarget{ref-SpecialIssue}{}%
{[}3{]} H. Sandberg S. Amin and K. H. Johansson, ``Special issue on
cyberphysical security in networked control systems,'' \emph{IEEE
Control Systems Magazine}, vol. 35, no. 1, 2015.

\leavevmode\hypertarget{ref-DistributedAlgorithms}{}%
{[}4{]} N. A. Lynch, \emph{Distributed Algorithms}. Morgan Kaufmann, 1996.

\leavevmode\hypertarget{ref-Kieckhafer1994}{}%
{[}5{]} R. M. Kieckhafer and M. H. Azadmanesh, ``Reaching approximate
agreement with mixed-mode faults,'' \emph{IEEE Transactions on Parallel
and Distributed Systems}, vol. 5, no. 1, pp. 53--63, 1994.

\leavevmode\hypertarget{ref-RobustNetwork}{}%
{[}6{]} H. J. LeBlanc, H. Zhang, X. Koutsoukos, and S. Sundaram,
``Resilient asymptotic consensus in robust networks,'' \emph{IEEE
Journal on Selected Areas in Communications}, vol. 31, no. 4, pp.
766--781, 2013.

\leavevmode\hypertarget{ref-Byzantine}{}%
{[}7{]} N. H. Vaidya, L. Tseng, and G. Liang, ``Iterative approximate
Byzantine consensus in arbitrary directed graphs,'' in \emph{Proc. 2012
ACM Symposium on Principles of Distributed Computing}, 2012, pp.
365--374.

\leavevmode\hypertarget{ref-RandomizedQuantize}{}%
{[}8{]} S. M. Dibaji, H. Ishii, and R. Tempo, ``Resilient randomized
quantized consensus,'' \emph{IEEE Transactions on Automatic Control},
vol. 63, no. 8, pp. 2508--2522, 2018.

\leavevmode\hypertarget{ref-msr}{}%
{[}9{]} S. M. Dibaji and H. Ishii, ``Resilient consensus of second-order
agent networks: Asynchronous update rules with delays,''
\emph{Automatica}, vol. 81, pp. 123--132, 2017.

\leavevmode\hypertarget{ref-LeBlanc2018}{}%
{[}10{]} H. J. LeBlanc and X. Koutsoukos, ``Resilient first-order
consensus and weakly stable, higher order synchronization of
continuous-time networked multiagent systems,'' \emph{IEEE Transactions
on Control of Network Systems}, vol. 5, no. 3, pp. 1219--1231, 2018.

\leavevmode\hypertarget{ref-Park2017}{}%
{[}11{]} H. Park and S. A. Hutchinson, ``Fault-tolerant rendezvous of
multirobot systems,'' \emph{IEEE Transactions on Robotics}, vol. 33, no.
3, pp. 565--582, 2017.

\leavevmode\hypertarget{ref-Kikuya2018}{}%
{[}12{]} Y. Kikuya, S. M. Dibaji, and H. Ishii, ``Fault-tolerant clock
synchronization over unreliable channels in wireless sensor networks,''
\emph{IEEE Transactions on Control of Network Systems}, vol. 5, no. 4,
pp. 1551--1562, 2018.

\leavevmode\hypertarget{ref-QuantizedController}{}%
{[}13{]} C. D. Persis, ``Robust stabilization of nonlinear systems by
quantized and ternary control,'' \emph{Systems \& Control Letters}, vol.
58, no. 8, pp. 602--608, 2009.

\leavevmode\hypertarget{ref-SelfTriggered}{}%
{[}14{]} C. De Persis and P. Frasca, ``Robust self-triggered
coordination with ternary controllers,'' \emph{IEEE Transactions on
Automatic Control}, vol. 58, no. 12, pp. 3024--3038, Dec. 2013.

\leavevmode\hypertarget{ref-EventTriggeredIntro}{}%
{[}15{]} W. P. M. H. Heemels, K. H. Johansson, and P. Tabuada, ``An
introduction to event-triggered and self-triggered control,'' in
\emph{Proc. 51st IEEE Conference on Decision and Control}, 2012, pp.
3270--3285.

\leavevmode\hypertarget{ref-DistEvent}{}%
{[}16{]} D. V. Dimarogonas, E. Frazzoli, and K. H. Johansson,
``Distributed event-triggered control for multi-agent systems,''
\emph{IEEE Transactions on Automatic Control}, vol. 57, no. 5, pp.
1291--1297, 2012.

\leavevmode\hypertarget{ref-Fan2015}{}%
{[}17{]} Y. Fan, L. Liu, G. Feng, and Y. Wang, ``Self-triggered
consensus for multi-agent systems with zeno-free triggers,'' \emph{IEEE
Transactions on Automatic Control}, vol. 60, no. 10, pp. 2779--2784,
2015.

\leavevmode\hypertarget{ref-Senejohnny2018}{}%
{[}18{]} D. Senejohnny, P. Tesi, and C. De Persis, ``A jamming-resilient
algorithm for self-triggered network coordination,'' \emph{IEEE
Transactions on Control of Network Systems}, vol. 5, no. 3, pp.
981--990, 2018.

\leavevmode\hypertarget{ref-EventMeanSquare}{}%
{[}19{]} L. Ma, Z. Wang, and H. Lam, ``Event-triggered mean-square
consensus control for time-varying stochastic multi-agent system with
sensor saturations,'' \emph{IEEE Transactions on Automatic Control},
vol. 62, no. 7, pp. 3524--3531, 2017.

\leavevmode\hypertarget{ref-MENGEventBased}{}%
{[}20{]} X. Meng and T. Chen, ``Event based agreement protocols for
multi-agent networks,'' \emph{Automatica}, vol. 49, no. 7, pp.
2125--2132, 2013.

\leavevmode\hypertarget{ref-StateConsensus}{}%
{[}21{]} F. Xiao and L. Wang, ``State consensus for multi-agent systems
with switching topologies and time-varying delays,'' \emph{International
Journal of Control}, vol. 79, no. 10, pp. 1277--1284, 2006.

\leavevmode\hypertarget{ref-Kashyap2007}{}%
{[}22{]} S. R. Kashyap A Basar T, ``Quantized consensus,''
\emph{Automatica}, vol. 43, no. 7, pp. 1192--1203, 2007.

\leavevmode\hypertarget{ref-Carli2010}{}%
{[}23{]} R. Carli, F. Fagnani, P. Frasca, and S. Zampieri, ``Gossip
consensus algorithms via quantized communication,'' \emph{Automatica},
vol. 46, no. 1, pp. 70--80, 2010.

\leavevmode\hypertarget{ref-Cai2011}{}%
{[}24{]} K. Cai and H. Ishii, ``Quantized consensus and averaging on
gossip digraphs,'' \emph{IEEE Transactions on Automatic Control}, vol.
56, no. 9, pp. 2087--2100, 2011.

\leavevmode\hypertarget{ref-EventTriggered}{}%
{[}25{]} Y. Wang and H. Ishii, ``Resilient consensus through event-based
communication,'' \emph{IEEE Transactions on Control of Network Systems},
vol. 7, no. 1, pp. 471--482, 2020.
\end{cslreferences}

\end{document}